\documentclass[superscriptaddress,rmp,showkeys]{revtex4}
\usepackage{amsmath}

\newcommand{\BiInfinity}	{ \overleftrightarrow {S} }
\newcommand{\biinfinity}	{ \overleftrightarrow {s} }
\newcommand{\Past}	{\overleftarrow{S}}

\newcommand{\past}	{\overleftarrow{s}}
\newcommand{\pastt}      {\overleftarrow{s}_{\!t}}
\newcommand{\pastprime}	{\past^{\prime}}
\newcommand{\Future}	{\overrightarrow{S}}
\newcommand{\future}	{\overrightarrow{s}}
\newcommand{\futuret}   {\overrightarrow{s}_{\!t}}

\newcommand{\CausalState}	{ {\cal S} }

\newcommand{\CausalStateSet}	{ \boldsymbol{\CausalState} }
\newcommand{\AlternateState}	{ {\cal R} }

\newcommand{\CausalEquivalence}	{ {\sim}_{\epsilon} }

\newcommand{\Prob}	{ {\rm Pr} }

\newcommand{\Cmu}	{ C_\mu }
\newcommand{\hmu}	{ h_\mu }
\newcommand{\EE}	{ {\bf E}}

\newcommand{\ObservedPartition}	{ \mathcal{F}}
\newcommand{\CausalPartition}	{\boldsymbol{\mathcal{G}}}
\newcommand{\MicroState}{\mathbf{Q}}
\newcommand{\microstate} {\mathbf{q}}
\newcommand{\TimeOp}{\mathrm{T}}

\newcommand{\FactorPartition}{\mathcal{Z}}
\newcommand{\FactorVariable}{Z}
\newcommand{\Observable}{S}
\newcommand{\observable}{s}
\newcommand{\observableprime}{\observable^\prime}
\newcommand{\ObservedStateSet}{\mathcal{X}}
\newcommand{\tprime}{t^\prime}

\newcommand{\eps}{\epsilon}

\newtheorem{proposition}{Proposition}

\begin{document}

\title{What Is a Macrostate? \\ 
Subjective Observations and Objective Dynamics}
\author{Cosma Rohilla Shalizi}
\email{cshalizi@umich.edu}
\affiliation{Center for the Study of Complex Systems, University of Michigan, Ann Arbor, MI 48109}
\affiliation{Santa Fe Institute, 1399 Hyde Park Road, Santa Fe, NM 87501}
\author{Cristopher Moore}
\email{moore@santafe.edu}
\affiliation{Computer Science Department, University of New Mexico, Albuquerque, NM 87131}
\affiliation{Physics and Astronomy Department, University of New Mexico, Albuquerque, NM 87131}
\affiliation{Santa Fe Institute, 1399 Hyde Park Road, Santa Fe, NM 87501}
\date{Begun 10 July 2000; last revised \today}

\begin{abstract}
We consider the question of whether thermodynamic macrostates are objective
consequences of dynamics, or subjective reflections of our ignorance of a
physical system.  We argue that they are both; more specifically, that the set
of macrostates forms the unique maximal partition of phase space which 1) is
consistent with our observations (a subjective fact about our ability to
observe the system) and 2) obeys a Markov process (an objective fact about the
system's dynamics).  We review the ideas of computational mechanics, an
information-theoretic method for finding optimal causal models of stochastic
processes, and argue that macrostates coincide with the ``causal states'' of
computational mechanics.  Defining a set of macrostates thus consists of an
inductive process where we start with a given set of observables, and then
refine our partition of phase space until we reach a set of states which
predict their own future, i.e.\ which are Markovian.  Macrostates arrived at in
this way are provably optimal statistical predictors of the future values of
our observables.

\end{abstract}

\maketitle

\tableofcontents

\section{What's Strange about Macrostates, or, Is It Just Me?}

Almost from the start of statistical mechanics, there has been a tension
between subjective or epistemic interpretations of entropy, and objective or
physical ones.  Many writers, for instance the late
E. T. \citet{Jaynes-essays}, have vigorously asserted that entropy is purely
subjective, a quantification of one's lack of knowledge of the molecular state
of a system.  It is hard to reconcile this story with the many physical
processes which are driven by entropy increase, or by competition between
maximizing two different kinds of entropy \citep{Fox-energy-evolution}.  These
processes either happen or they don't, and observers, knowledgeable or
otherwise, seem completely irrelevant.  In a nutshell, the epistemic view of
entropy says that an ice-cube melts when I become sufficiently ignorant of it,
which is absurd.

These difficulties with entropy are only starker versions of the difficulties
afflicting all thermodynamic macroscopic variables.  Their interpretation
oscillates between a purely epistemic one (they are the variables which we
happen to be willing and able to observe) and a purely physical one (they have
their own dynamics and have brute physical consequences, e.g.\ for the amount
of work which engines can do).  These difficulties are inherited by our
definition of macrostates.  Standard references define macrostates either as
sets of microstates, i.e.\ subsets of phase space, with given values of a small
number of macroscopic observables
\citep{Landau-Lifshitz,Reichl-modern,Baierlein-thermal-physics}, or probability
distributions over these
\citep{Ruelle-stat-mech-rigor,Balian-from-micro-to-macro}.  A given set of
observables induces a set of macrostates, which form a partition of the phase
space;\footnote{\citet[p.~77n]{Tolman-stat-mech}, while objecting to the {\em
    name} ``macrostate,'' regards it as a partition, as above.}  but why is one
such partition better than another?

There are generally several different sets of macroscopic variables which can
be observed a given system.  In some cases, different sets of observables are
equivalent, in the sense that they induce the same partition of the phase
space, and so their macrostates are in one-to-one correspondence; for instance,
for an ideal gas with a constant number of molecules, we obtain the same
macrostates by measuring either pressure and volume, or temperature and
entropy.  In other cases, observing different sets of variables will partition
the set of microstates in different ways --- producing partitions that are
finer, coarser, or incomparable.

Even if we restrict our attention to extensive variables, there is a hierarchy
of increasingly disaggregated, fine-grained levels of description, with
associated macroscopic variables at each level.  At the highest and coarsest
level are thermodynamic descriptions, in terms of system-wide extensive
variables or bulk averages.  Below them are hydrodynamic descriptions, in terms
of local densities of extensive quantities.  Below them is the Boltzmannian
level, described with occupation numbers in cells of single-molecule phase
space, or, in the limit, phase-space densities.  (Below the Boltzmannian level
we get densities over the whole-system phase space, and so statistical
mechanics proper.)  We can sometimes demonstrate, and in general believe, that
we can obtain the coarser descriptions from the finer ones by integration or
``contraction'' \citep[cf.][ch.~9]{Keizer-stat-thermo}.  Thus there are many
hydrodynamic macrostates for a given thermodynamic one, i.e.\ the hydrodynamic
partition is much finer.

Clearly there is a problem here if macrostates are {\em purely} objective.  In
that case, we should be forced to use {\em one} level of description.  On the
other hand, we can formulate and test theories at all levels of description,
and we know that, for instance, both thermodynamic and hydrodynamic theories
are well-validated for many systems.

We hope to offer a resolution along the following lines.  Intelligent creatures
(such as, to a small extent, ourselves) start with certain variables which they
are able to observe, and which interest them.  This collection of variables
defines a partition of phase space.  This partition may not be an optimal
predictor of its own future; it may have non-Markovian dynamics, with
unaccounted-for patterns in its variables' time series.  In such cases,
intelligent observers postulate additional variables, attempt to develop
instruments capable of observing them, and thus refine this partition.  Our
proposal is that good macrostates are precisely the partitions at which this
process terminates, i.e.\ refinements of the observational states whose
dynamics are Markovian.  One can show that there is a unique coarsest such
refinement, given the initial set of observables, and that this refinement is
provably optimal in several senses as a statistical predictor of the future.
To prove these things we must give a brief summary of the theory of the causal
architecture of stochastic processes, known as computational mechanics.

Definitions of Markov processes and some basic information theory can be found
in the Appendix.

\section{Computational Mechanics and Causal States}

In the late 1970s and early 1980s, workers in nonlinear dynamics developed
methods, called ``attractor reconstruction'' or ``geometry from time series,''
to reconstruct the vector field of a dynamical system from time series of
measurements of some function of the state
\citep*{Geometry-from-a-time-series,Takens-embedding,Kantz-Schreiber}.
Inspired by this, since 1989 Crutchfield et al.\
\citep{Inferring-stat-compl,JPC-DPF-stat-compl-of-1d-spin,TDCS,DNCO,CMPPSS}
have formulated a theory, ``computational mechanics,'' which constructs, from
observations of a stochastic process, the minimal model capable of generating
that process.  Put differently, they have developed a technique for discovering
and representing all the predictive patterns in a time series.

The minimal model produced by computational mechanics represents the causal
architecture of the process, or alternately how it stores and processes
information (hence the term {\em computational} mechanics).  If statistical
mechanics is a ``forward'' approach, deriving macro-consequences of
micro-dynamics, computational mechanics is an ``inverse'' approach, finding
minimal causal architectures capable of producing the statistics of observed
time series.

The key notion in computational mechanics is that of {\em causal state}, which
works like this.  We observe a stochastic process, which we break at arbitrary
points into ``histories'' and ``futures.''  Two histories belong to the same
causal state if and only if they are equivalent for predicting the future,
i.e., if they lead to the same conditional probability distribution for
futures.

Formally, consider a discrete-time stochastic process, stretching to infinity
in both directions: $\ldots s_{t-1}, s_t, s_{t+1} \ldots = \biinfinity$.  Break
the process into two parts: one, the ``past'' or ``history,'' is all the values
up to and including time $t$, which we write $\pastt$; the other, the
``future'' $\futuret$, all the values after that time.  We write $\Past$ and
$\Future$ for the set of all possible histories and futures respectively.  If
the system is in equilibrium and this time series is stationary, we can drop
this subscript, and we assume this for now.  We wish to predict $\future$ on
the basis of $\past$.

Now, any prediction method treats some histories as equivalent to each other;
for instance, if we model the system as only depending on its $k$ previous
values, two histories which last differed $k$ time-steps ago are equivalent.
Moreover, all we need to predict about the future is which equivalence class we
will find ourselves in.  Thus any prediction method induces a partition over
the phase space, and for any partition there is some probability distribution
over the future equivalence classes given which one we are in now.

There are many ways to construct partitions about which we can make correct
predictions.  For instance, we could lump all of phase space into a single,
trivial ``macrostate,'' and announce that given we are in that state now, we
will be in the future as well.  But obviously such a scheme fails to capture
anything important about the system.  Computational mechanics seeks a partition
which is as coarse as possible, but which captures all available information
about the time series of observables.  That is, it gives optimal predictions of
our observables while attributing as little structure to the system as
possible.

With this in mind, we claim that the optimal partition is simply the following.
Say that two histories $\past, \pastprime$ are {\em causally equivalent} if and
only if they give the same conditional distribution for futures:
\[
\past \CausalEquivalence \pastprime \mbox{ iff }
\Prob[\future | \past] = \Prob[\future|\pastprime]
\]
This relation $\CausalEquivalence$ is symmetric, reflexive and transitive, and
thus divides the set $\Past$ of all pasts into equivalence classes.  We define
the {\em causal state} of a history as its equivalence class under
$\CausalEquivalence$:
\[
\eps(\past) \;\equiv\; \left\{
\pastprime \,|\, \pastprime \CausalEquivalence \past \right\}
\]
Then it is clear that the dependence of the system on its past is completely
captured by its causal state,
\[ \Prob\left[ s_{t+1} | \pastt \right]
 \;=\; \Prob\left[ s_{t+1} | \eps(\pastt) \right]
\]

We now formalize our claim that the causal states form an optimal partition of
histories.  For a given partition $\AlternateState$ of histories into
equivalence classes, we consider the mutual information
$I[\future;\AlternateState]$ between it and the system's future (for
definitions of information-theoretic quantities, see Appendix
\ref{app:info-theory}).  This quantity is limited by the mutual information
between the system's past and its future, $I[\future;\AlternateState] \leq
I[\future;\past]$.  We call partitions which attain this limit {\em prescient}.
\citet{CMPPSS} showed that the causal states are prescient.  Moreover, if two
histories are equivalent with respect to some prescient partition, they are
almost always in the same causal state as well.  Thus, except for a set of
measure zero, any prescient partition is a refinement of the causal states, so
the causal states form the coarsest possible prescient partition.

Moreover, the causal states are the {\em least complex} prescient partition in
the following sense.  Given a partition into equivalence classes
$\AlternateState$ we define its {\em statistical complexity} as the entropy
$H[\AlternateState]$.  The statistical complexity is the amount of information
which the partition encodes about the past.  Normally in statistical mechanics
we seek to {\em maximize} the entropy, but here entropy measures, not the
unbiasedness of a distribution, but the complexity of a predictor, and so in
minimizing it we are applying Occam's Razor.  We can therefore say that the
optimal predictor is the prescient predictor of minimal statistical complexity.
In fact, for all prescient partitions $\AlternateState$ we have
$H[\AlternateState] \geq H[\eps(\past)]$, so the causal states are the unique
prescient states of minimal statistical complexity (\citet{CMPPSS}, App.\ D of
\citep{CRS-thesis}).

Because the causal states are optimal predictors, we can say that the
statistical complexity of the process is equal to their statistical complexity;
we denote this $\Cmu = H[\eps(\past)]$.  The statistical complexity of the
process is therefore the amount of information about its past which is {\em
  relevant} to its future, the amount retained internally, so to speak.  It
also has a nice physical interpretation connected to thermodynamic entropy,
which we give in the next section.

A common measure of the complexity of a stochastic process, and of the amount
of information stored in it, is the predictive information
\citep*{Bialek-Nemenman-Tishby}, effective measure complexity
\citep{Grassberger-1986} or ``stored information'' \citep{Shaw-dripping}, $\EE
\equiv I[\future;\past]$.  Since the causal states are prescient,
$I[\future;\past] = I[\future;\eps(\past)] \leq H[\eps(\past)]$ (see
Appendix~\ref{app:info-theory}), so $\Cmu \geq \EE$.  While a complete
knowledge of the causal state reduces the uncertainty in the future by $\EE$
bits, one needs $\Cmu$ bits to make this prediction.  In general
\citep{CMPPSS}, $\EE \leq \Cmu$, and for the specific case of first-order
Markov processes, $\Cmu = \EE + \hmu$, where $\hmu \equiv H[s_{t+1}|\pastt]$ is
the {\em entropy rate} of the process, also equal to $H[s_{t+1}|\eps(\pastt)]$.
These notions will be useful later on, when we propose a definition of
emergence based on the ratio between predictive information and statistical
complexity.

Readers familiar with the literature on statistical explanation will recognize
that the the causal state partition can also be though of as an application of
Salmon's notion of a ``statistical relevance basis'' to stochastic processes
\citep{Salmon-1971,Salmon-1984}.

If we consider the time series of causal states $\ldots, \eps(s_{t-1}),
\eps(s_t), \eps(s_{t+1}), \ldots$, we have a new stochastic process.  Moreover,
since each causal state $\eps(s_t)$ contains all the relevant information about
its entire past $\pastt$, this is a Markov process, i.e.\ the probability
distribution of futures depends only on the current state (see Appendix
\ref{app:markov}).  Thus we have collapsed the original process, regardless of
its dependence on its history, into a Markov process --- but one which contains
all relevant information about the original process.

The observed process is a random function of this Markov process, i.e., a kind
of ``hidden Markov model''.  The Markov properties of the causal states justify
in part their name, since they are exactly the ``screening-off'' properties
that have long been recognized as essential to causation \citep{Salmon-1984},
and which form the basis of statistical methods of causal inference for
non-dynamical systems \citep*{Spirtes-Glymour-Scheines,Pearl-causality}.

If the underlying process $\BiInfinity$ is not stationary, all is not lost.
Formally, in fact, we can generalize the theory to arbitrary processes.  For
our purposes here, however, we only need the idea of a conditionally stationary
process, which is to say one in which $\Prob(\futuret|\pastt=\past) =
\Prob(\future_0|\past_0 = \past)$, for all times $t$ and histories $\past$.
The above theory then carries over directly (stationary processes are all
conditionally stationary as well), with the exception that the probability
distribution of the causal states, and $H[\CausalState]$, can be a function of
time.

We have spoken throughout as we knew all the necessary conditional
probabilities exactly.  This is sometimes the case with analytical models
\citep{DNCO}, but never with experimental data.  However, there are {\em
  reconstruction} algorithms which, under mild statistical assumptions, will
converge to the correct causal states, given sufficient experimental data
\citep{AfPDiTS}.  For our present purposes, it is enough to know that causal
states can be inferred reliably from observations.

\section{Causal States from Coarse-Grained Observations}

Consider our favorite statistical-mechanical system.  It has a phase space
$\Gamma$, every point $\microstate$ of which is a complete specification of the
positions, momenta, spins, etc. of all particles.  Discretizing time, as is
common, we say that the evolution on $\Gamma$ is governed by an operator
$\TimeOp$: $\microstate_{t+1} = \TimeOp \microstate_{t}$.  We do not rule out
the possibility that $\mathrm{T}$ is stochastic, but we insist that there are
no ``hidden variables,'' so that $\left\{\microstate_{t}\right\}$ form a Markov
chain.  We are concerned with an ensemble of such systems, so we write the
random variable for the current microstate $\MicroState$.  We do {\em not}
assume that the ensemble is any of the usual thermodynamic ensembles, or even
that the distribution of $\MicroState$ is invariant.

The system changes over time.  We probe it with observations of limited
precision at each time step.  What our probes give us are many-one functions of
the location in phase space.  More formally, the observation process is
represented by a function $f: \Gamma \mapsto \Xi$, where $\Xi$ is our favorite
(possibly multi-dimensional) space for representing
observations\footnote{Strictly, $f$ should map to a distribution of observed
  values, to represent fluctuations and noise, in which case macroscopic states
  would be defined by probability distributions of macroscopic variables.
  This, however, would increase the complexity of our exposition without a
  corresponding gain in insight, so we will pretend that our observations are
  exact and noiseless.}.  This function $f$ partitions the phase space
$\Gamma$, i.e., it divides it into mutually exclusive and jointly exhaustive
sets, on each of which $f$ takes a unique value.  Let the partition of $\Gamma$
induced by $f$ be $\ObservedPartition$.  Then $f(\MicroState) = \Observable_t$
defines another process, which need not be Markovian.  Call it the {\em
  observed process}.

We now form the causal states of the observed process.  Our observations, as
noted, induce a partition of the phase space.  Therefore a sequence of
observations induces a refinement of that partition.  Each observation value
$x$ corresponds to a set $\ObservedPartition_x$ of points in phase space.  The
sequence of observations $x,y$ thus corresponds to the set
$\ObservedPartition_{x,y} \equiv \ObservedPartition_y \cap \TimeOp
\ObservedPartition_x$, i.e., those points at which we observe $y$ now, and at
which we would have observed $x$ one step back.  The sets
$\ObservedPartition_{x,y}$ are a refinement of the observed partition
$\ObservedPartition$, and we can extend this to countable sequences of
observations.  The set of causal states, $\CausalStateSet$, is a partition on
the set $\Past$ of observational histories.  Therefore it induces a partition
on $\Gamma$ which is a coarsening of the partition induced by infinite-length
histories.  Call this partition $\CausalPartition$.  Each causal state
therefore corresponds to a region of phase space, which in principle is
accessible to some coarse-grained observational procedure.  Observations of
this variable form a new stochastic process $\left\{\CausalState_t\right\}$
which is Markovian, and a knowledge of $\CausalState_t$ is all that is needed
to predict $\futuret$ optimally.

Let us write the partition on $\Past$ induced by the present observation as
$\ObservedStateSet$.  What is the relationship between the causal partitions,
$\CausalPartition$ and $\CausalStateSet$, and the corresponding observational
partitions, $\ObservedPartition$ and $\ObservedStateSet$?  There are four
possibilities:
\begin{enumerate}
\item The observational and the causal partitions are the same.
\item The causal partition is a refinements of the observational one.
\item The causal partition is coarser than the observational one.
\item The causal and observational partitions are incomparable.
\end{enumerate}
In the next sections we explain the physical meaning of cases (1--3), and give
physical examples of them.

\subsection{The Observables Define a Macrostate}

Suppose two observational histories, $\ldots{\observable}_{t-2}
{\observable}_{t-1} {\observable}_{t}$ and
$\ldots{\observableprime}_{{\tprime}-2} {\observableprime}_{{\tprime}-1}
{\observableprime}_{{\tprime}}$ are causally equivalent iff ${\observable}_{t}
= {\observableprime}_{{\tprime}}$.  This means that the current causal state is
defined by the current values of the macroscopic observables, and conversely
any difference in a macroscopic observable means a difference in causal state.
In the notation introduced earlier, $\CausalStateSet = \ObservedStateSet$ iff
$\ObservedPartition = \CausalPartition$.  The macrostates then have all the
properties of causal states.  Their dynamics are Markovian and statistically
reproducible, and no prediction of future values of the macrovariables can be
better than one based simply on their present value.  An obvious example is the
combination of pressure, volume and temperature for an ideal gas near
equilibrium.

In such cases, we can give a nice interpretation to the statistical complexity
$\Cmu$.  Recall that $\Cmu = H[\CausalState]$, the amount of information needed
to specify the causal state.  Because the causal state and the macrostate
$\Observable$ are equivalent, $H[\CausalState] = H[\Observable]$.  But
$\Observable = f(\MicroState)$, so $H[\Observable|\MicroState] = 0$ --- if we
knew the exact microstate, there would be no uncertainty in the macrostate.
Now, for any two random variables, $H[X,Y] = H[X] + H[Y|X]$.  Let us make both
possible decompositions of $H[\MicroState,\Observable]$.
\begin{eqnarray*}
H[\MicroState|\Observable] + H[\Observable] & = & H[\Observable|\MicroState] +
H[\MicroState] \\
H[\MicroState|\Observable] + \Cmu & = & H[\MicroState] \\
\Cmu & = & H[\MicroState] - H[\MicroState|\Observable] \\
\Cmu & = & I[\MicroState;\Observable]
\end{eqnarray*}
That is, the statistical complexity is just the amount of information about the
microstate that is contained in the macrovariables.

Since the macrostates form a first-order Markov chain, there is, as we
mentioned above, a simple relationship between the statistical complexity, the
entropy rate, and the predictive information, viz., $\EE = \Cmu - \hmu$.  Since
$\EE = I[\future;\CausalState]$, and $\hmu = H[\Observable_1|\past]$, we have
$I[\future;\CausalState] = H[\CausalState] - H[\future^1|\CausalState]$.

\subsection{The Causal States Are Finer than the Macrostates}

Suppose $\CausalPartition$ is a refinement of $\ObservedPartition$, or,
equivalently, $\CausalStateSet$ is a refinement of $\ObservedStateSet$.  Then,
in addition to knowing the current values of the macrovariables, we must know
something of their history as well.  Or, more exactly, if we do not, we do not
have a causally complete set of macrovariables, and the observed dynamics are
not only non-Markovian, they cannot be optimally predicted.  However, they {\em
  can} be optimally predicted from a knowledge of $\CausalState$, whose
time-evolution {\em is} Markovian.  Moreover, if we know what cell of
$\CausalPartition$ the system is in, we know the value of $\CausalState$, which
suggests that, in principle, there is a observational procedure which will tell
us how to optimally predict our original macrovariables.

We can go one step further, however, by invoking a result about refinements of
partitions (see the appendix).  Suppose $A$ is a partition and $B$ is a
refinement of it.  Then there exists at least one {\em minimal factor
  partition} $C$ such that $B$ is the product of $A$ and $C$, and this is not
true of any partition with fewer cells than $C$.  Since $\CausalPartition$ is a
refinement of $\ObservedPartition$, there is therefore at least one
$\FactorPartition$ such that $\CausalPartition = \ObservedPartition \cdot
\FactorPartition$.  If we observe the macrovariable $\FactorVariable$
corresponding to $\FactorPartition$ together with our original macrovariables,
it is the same as if we had observed the causal state directly, and so we get a
causally complete set of macrovariables and nice macrostates.  In other words,
we if observe either $\CausalState$ or
$\left(\Observable,\FactorVariable\right)$, we reduce the present case to that
in the previous subsection.

Generally, there are a very large number of minimal factor partitions which can
take the role of $\FactorPartition$.  Which one we observe is dictated by
practical considerations --- experimental accessibility, smoothness of the
resulting macrovariable over phase space, degree of uncertainty in the
macrovariable, etc.  This should not be worrisome, however, since there are
elementary cases where we can complete a set of macrovariables in more than one
way.  Given pressure and volume for an ideal gas, for instance, we get the same
macrostates from observations of temperature or molecule number.

It is worth noting that the minimal factor partition $\FactorPartition$ is
incomparable to $\ObservedPartition$, and so in some sense orthogonal or
unpredictable from $\ObservedPartition$.  Clearly, there is a bijection between
$\CausalState$ and $\left(\Observable,\FactorVariable\right)$.  Hence
$H[\CausalState] = H[\Observable,\FactorVariable] =
H[\FactorVariable|\Observable] + H[\Observable]$.  Since $H[\Observable]$ does
not depend on our choice of factor variable $\FactorVariable$, it follows that
$H[\FactorVariable|\Observable]$ is the same for all factor variables, i.e.,
they all have the same degree of uncertainty remaining once we know the
original observables.  Furthermore, all the information they contain is
relevant to the causal state:
\begin{eqnarray*}
I[\CausalState;\FactorVariable] & = & H[\CausalState] - H[\CausalState|\FactorVariable] \\
& = & H[\Observable,\FactorVariable] - H[\Observable,\FactorVariable|\FactorVariable]\\
& = & H[\Observable,\FactorVariable] - (H[\Observable,\FactorVariable,\FactorVariable] - H[\FactorVariable])\\
& = & H[\Observable,\FactorVariable] - H[\Observable,\FactorVariable] + H[\FactorVariable]\\
& = & H[\FactorVariable]
\end{eqnarray*}
Similarly, $I[\CausalState;\FactorVariable|\Observable] =
H[\FactorVariable|\Observable]$, which is independent of the factor partition
we use.

Sadly, there is no guarantee that any of the factor partitions are
experimentally accessible, still less accessible by practical or easy
experimental procedures.  In such cases, however, we may still eliminate memory
effects from our models by constructing the causal states from observational
histories.\footnote{For more on reducing dialectical or historical explanations
  to mechanical ones via causal states, see \citet{CRS-thesis}.}

For an example of this method in (unwitting) action, consider hysteresis in
ferromagnets.  The response of a ferromagnetic substance to a magnetic field
can be treated, equivalently, either as a function of its past history of
applied fields, or as a function of the current applied field and the
magnetization.  Another example is provided by the study of chaotic dispersion
in fluids jets \citep{Cenci-et-al-mixing-in-jet}.  The initial measurement
partition here involves the character of the motion of the jet, and shows
strong memory effects, significantly complicating the analysis.  Recent work
has shown how to eliminate these memory effects, by refining the partition of
the state space in just the way we suggest above
\citep{Lacorata-et-al-Markov-for-long-memory,%
  Abel-et-al-hierarchical-Markovian-modeling}.

\subsubsection{An Apparent Counterexample: Disordered Materials}

Amorphous solids \citep{Zallen-amorphous} and their magnetic equivalents, spin
glasses \citep{Fischer-and-Hertz-spin-glasses} are remarkable not just because
of they display slow dynamics, but because they display distinct dynamics on an
immense range of time-scales.  A crude but graphic illustration is given by
ordinary silicate glass.  Under mechanical stresses with short characteristic
times, it is brittle; under stresses with long characteristic times, it is
effectively liquid.  Spin glasses, similarly, can display distinct
susceptibilities to oscillatory magnetic fields over sixteen orders of
magnitude in frequency \citep{Fischer-and-Hertz-spin-glasses}.  Since the two
cases are basically similar, but the requisite physical theory is easier to
grasp for spin glasses, we will concentrate on them.

This hierarchy of time-scales implies that memory effects are very important in
disordered materials.  Indeed, many of the usual assumptions made in
discussions of statistical mechanics, such as having an ``aged'' ensemble at
equilibrium, are simply nonsensical in these cases.  At low temperatures, the
slowest time-scales can be geologically significant.  To work with samples
which have aged into equilibrium requires literally inhuman longevity (to say
nothing of patience).  While technologies which would allow this have been
proposed \citep{Dyson-time-without-end}, they are not yet common in
laboratories.  Have we here found substances where our approach to eliminating
memory effects breaks down?  And what does one do, if one cannot use properly
aged and equilibrated ensembles?

The physical mechanism responsible for the long time-scales actually holds the
answer to both questions.  Each spin in a spin glass participates in a mixture
of ferromagnetic and antiferromagnetic interactions of varying strengths.  The
result is generally frustration, i.e., no setting of the spins minimizes all
interaction energies simultaneously.  This leads to the existence of numerous
local minima in the energy landscape, generally with widely varying energies,
and so widely varying heights of the barriers separating them.  One must either
flip many spins at once, or equivalently make many energetically-unfavorable
spin flips in succession, to get from one minimum to another.  The time it
takes to pass between minima will generally be exponential in the height of the
energy barrier between them, as one expects from the Arrhenius equation.  (The
causes and details of frustration in glass are different, but the overall
picture is similar.)  Thus, on a given time-scale, barriers above a certain
height are effectively infinite, i.e., there probability of crossing them is
negligible.  The spin glass is thus effectively confined to a fixed region of
phase space.  Within this region, the local minima define metastable states,
with characteristic life-spans, and so relative probabilities, that reflect the
heights of the barriers surrounding them.

We can thus see the way to eliminating memory effects: one takes as one's
macrovariables the occupancy probabilities of the metastable local minima.
Those in the effectively-inaccessible region do not contribute.  Within the
accessible region of phase space, there is a more-or-less gradual leakage of
probability from the initial metastable state to the others.  To extrapolate
this forward, however, we do not need to know the history of that seepage,
merely the current distribution over the local minima.  In fact, this is a
common theoretical ploy, sometimes spoken of as employing ``a macroscopic
number of macroscopic degrees of freedom''
\citep{Fischer-and-Hertz-spin-glasses}.  Experimentally, one never studies an
equilibrium ensemble, but rather one that is always aging, and it is precisely
the aging properties which are of interest!

\subsection{The Macrostates Are Finer than the Causal States}

Suppose $\ObservedStateSet$ is a refinement of $\CausalStateSet$.  Then some
distinct values of the macrovariables have exactly the same consequences for
the future evolution of the macrovariables.  The distinction between those
macrostates is meaningless, and some of the details in those macrostates is
superfluous.  There are several reasons, by no means mutually exclusive, why
this might be so.

First, some of our variables could be irrelevant, given the others.  More
precisely, future events could be statistically independent of the value of
variable $Y$ given the present value of other variables $X$.  It is hard to
find examples of this in statistical mechanics proper, simply because those
variables have been subject to a long process of (informal) selection for
relevance, but it is easy to find examples of this in other domains of
scientific inquiry.  Techniques for identifying, or constructing, combinations
of variables which render others irrelevant play a major role in statistical
methods of causal inference \citep*{Spirtes-Glymour-Scheines,Pearl-causality}.
(Note that if one macrovariable is a deterministic function of the others, then
we get the same partition of $\Gamma$ whether or not we adjoin it to the
others.  Similarly, the partition of histories we get is the same.)

Second, our observational procedure could encode an ``unphysical'' distinction.
In nematic liquid crystals, for instance, an important role is played by the
``director'', a local vector indicating the average direction of orientation of
the rod-shaped molecules in the neighborhood of a point.  However, the
molecules in a nematic are symmetric when their long axis is inverted, so the
director is not a normal vector, but one in which opposite vectors are
identified, i.e., $\mathbf{n} = - \mathbf{n}$.
\citep{Collings-liquid-crystals,de-Gennes-Prost-liquid-crystals}.  If we did
not know this, however, and tried to observe the director as an ordinary
vector, we would find that which of two opposite observations we got for the
director would be a matter of pure chance, i.e., an artifact, and that we would
retain full predictive power if we identified opposite director vectors, i.e.,
if we coarsened our observational partition.

Finally, we may have an unpredictable variable, in the following sense.  On the
one hand, it takes on significantly different values in regions of the phase
space which are visited under the dynamics.  On the other hand, given the time
scale separating our observations, the dynamics randomizes those values so
thoroughly that little or no effective prediction of the variable is possible.
In these cases, the variable ``washes out'' from the partition which maximizes
predictive power, namely $\CausalPartition$.  In extreme cases, none of the
variables has any predictive power, at the time-scale and resolution available
to us, so the observed process becomes a sequence of IID random variables, and
$\CausalPartition$ becomes the trivial partition on $\Gamma$.  For example,
consider a liter of ideal gas at standard temperature and fixed, normal
molecule number.  If we observe pressure and internal energy (to reasonable
precision) at intervals of one year, the dynamics will have so thoroughly mixed
phase space that our original observations will have absolutely no predictive
value at all.  \citep[cf.][ch.~12]{CRS-thesis}.  In such cases, there is simply
no point in making predictions, and one's resources are better used
elsewhere.\footnote{In the real world, it is often not obvious when a variable
  contains predictive information, at least at the time- and resolution- scale
  of interest, and great efforts can be devoted to ever-more-elaborate
  deterministic models of what are, to all intents and purposes, coin-flips.
  For an example of causal state reconstruction showing that some variables
  contained no useful information, and how recognition of this led to improve
  predictions, see \citet*{Palmer-complexity-in-atmo}.  Of course, variables
  which are effectively IID over long times or at coarse resolution can contain
  a lot of predictive information at finer scales.  We will return to this
  point later; cf.\ the hierarchical scaling complexities of
  \citet{Badii-Politi}.}

\subsection{The Physical Meaning of the Causal States}

Starting from the observational variables, one can construct the the causal
states, and from them the minimal coarse-grained observation which allows for
optimal prediction of the original observables.  If the two do not coincide,
one can profitably replace the original set of observables with a new one,
either by adjoining new observations or by eliminating unphysical distinctions
or variables without predictive power.  In any case, we can construct, from the
original macrovariables, a new set of macrovariables whose macrostates are
their own causal states.

These well-constructed macrostates have a number of properties it is worth
noting.  First, their statistical complexity is just the amount of information
the macrovariables contain about the microstate --- how much our uncertainty
about the microstate is reduced by learning the macrostate.  Second, the
macrostates are Markovian.  This means that they will be mixing just when they
satisfy the conditions for Markov processes to be mixing.  This can be true
even when $\TimeOp$ is {\em not} mixing.  Third, again because the macrostates
are Markovian, there is a Gibbs distribution over sequences of macrostates
\citep{Bremaud-markov-chains}.  We have not had to assume any sort of
equilibrium property, however, and this may be part of the reason why Gibbs
distributions are still useful out of equilibrium.\footnote{We owe this last
  suggestion to conversation with Erik van Nimwegen, but are pretty sure he
  disagrees.}

We began with certain arbitrary or subjective decisions, about which variables
to observe --- about what partition $\ObservedPartition$ of $\Gamma$ to employ.
Our desire to have dynamics with good causal properties (Markovianity, etc.)
led us first to refine that partition by considering observational histories,
and then to group together histories in constructing the causal states.
Whether, at that point, we end up adjoining new variables to our original
macrovariables, leaving them alone, or even coarsening them, has nothing to do
with our experimental decisions or epistemic hankerings, merely the purely
mechanical, physical, objective microdynamics.  In the causal states we have
arrived, so to speak, at objective explanations of subjective quantities.
\citep[cf.][]{TDCS}

\section{Levels of Description and Emergence}

Earlier, we raised the puzzle of how different levels of description of the
same system can co-exist.  The answer, we propose, is that different causal
states are induced by different measurement partitions.  Consider two
measurement partitions, one a coarsening of the other.  The coarser measurement
is therefore a function of the finer one, and is on a higher, less specific
level of description.  Suppose the finer, lower-level measurement partition is
causal.  It is well-known that the Markov property does not generally survive
coarsening the states, which means that its coarse-grainings will not, in
general, be their own causal states.  The causal states of the coarse-grained
measurements are well-defined, however, and cannot be any finer than the
states of the fine-grained measurement partition.  It is possible, however,
that one does not need to go all the way back to the original partition
to find those causal states --- in fact, there is no reason the coarse-grained
measurements cannot be identical to their own causal states.  We then have
two levels of description, and can give a coherent causal
account at each level.

In this section, we explore the uses of these ideas in interpreting statistical
mechanics, and suggest a definition of emergence.  We start by considering the
old question of the relationship between molecular dynamics and thermodynamics,
in the particularly transparent context of the fluctuations of a gas at
equilibrium.  This leads us to suggest a definition of ``emergence''.  We then
clarify the relationship between generalized hydrodynamics and thermodynamics,
and attempt to explain the ubiquity of Gibbs distributions for macroscopic
configurations.  Finally, we look at the practice of cellular automata and
lattice gas modeling for examples of deliberately constructing adequate
coarse-grainings.

\subsection{Equilibrium Fluctuations and a Definition of Emergence}

Systems prepared ``in equilibrium'' actually fluctuate continually.  If our
observations are sufficiently coarse, then we will essentially only see
fluctuations about equilibrium which leave us in the linear regime.  In that
case, the Onsager theory provides the tools to describe the fluctuations, and
to do so in terms of the same variables which work at equilibrium
\citep{Keizer-stat-thermo}.

Consider a cubic centimeter of argon at fixed temperature, pressure and number.
(For a more detailed version of what follows, see \citet[sec.\
  11.2.3]{CRS-thesis}).  The only macrovariable left to fluctuate is the
internal energy.  One can calculate, from the Onsager theory, that the Shannon
entropy of the internal energy is 33.3 bits.  Taking a time-step of one
millisecond, the entropy rate, i.e., the rate at which the uncertainty
increases, is 4.4 bits.  The predictive information is thus $33.3 - 4.4 = 28.9$
bits.  In doing the corresponding calculations for the microstates, we start
with the fact that the microstate is its own causal state, since (almost by
definition) it is Markovian.  Thus $\Cmu = 6.6 \cdot 10^{20}$ bits.  If we take
the time-step to be one nanosecond, one can estimate $\hmu$
\citep{Gaspard-chaos-scattering-stat-mech} to be $3.3 \cdot 10^{20}$ bits, with
$\EE = 3.3 \cdot 10^{20}$ bits.

Following \citet{Palmer-efficiency}, we define the predictive efficiency of a
process as the fraction of the information it contains which actually effects
the future, i.e., as the ratio $\EE/\Cmu$.  We then see that the macrostates
can be predicted with much higher efficiency (0.87) than the microstates (0.5).
Indeed, this comparison is rather unfair to the macrostates, since we are
predicting them over a much longer time-scale.  If we predicted them at the
same time resolution as the microstates, we would find that the efficiency of
prediction was essentially one.  Conversely, if we tried to predict the
microstates at the macro time-scale, we would find an efficiency of prediction
of essentially zero.  Yet the macrovariables are transparently a function, a
coarse-graining, of the microvariables.

This leads us to define a relation of ``emergence'' between two sets of causal
variables if (1) one is a coarse-graining of the other and (2) the
coarse-grained variables can be predicted more efficiently.  In this sense, we
can be precise about the long-standing intuition that thermodynamics emerges
from statistical mechanics: thermodynamic variables are more informative about
their own dynamics.  This also gives us a hint as to what constitutes a {\em
  good} set of macrovariables: it should not just be causally complete, but
also more predictively efficient than the microvariables.  This is not always
the case; sometimes coarse-grainings are less efficiently predictable than the
original variable, a condition which J. P. Crutchfield (personal communication)
has designated ``submergence''.

\subsection{Hydrodynamics and Levels of Description}

One of the more important developments in statistical mechanics and condensed
matter physics has been the rise of ``generalized hydrodynamics,'' where
description centers on the local densities of extensive quantities and order
parameters \citep{Forster-hydro,Chaikin-Lubensky}.  Normal hydrodynamics is
included as a special case.  We are not going to expound this theory,
interesting though it is.  Rather, we wish to draw out two points.

The first is that many (perhaps all) systems which are adequately described at
the hydrodynamic level can also be described, accurately but less precisely, at
the thermodynamic level.  This is perfectly sensible from our point of view.
If one starts with observations of local densities, it is extremely unlikely
that these will be adequately predicted from purely global quantities.  The
causal states one forms remain, therefore, tied to local densities.
Conversely, knowledge of the local densities is excessive if all you want to
predict are their global averages or sums.  The two descriptions coexist,
because they are intended to answer different questions --- not because one is
more objective than the other.

Second, one can show \citep*{CRS-thesis} that the relationship between
the hydrodynamic description and the thermodynamic one is generally one of
emergence, in the sense described above.  This is comforting, since one can
generally ``contract'' hydrodynamic descriptions into thermodynamic ones
\citep{Keizer-stat-thermo}.  Similarly, the hydrodynamic level itself emerges
from the thermodynamic one.  Third, when one constructs local causal states
\citep*[following][]{CRS-thesis}, one finds that they generally form a
Markov random field.\footnote{No counter-examples are known.  Whether they
  always form a Markov random field is currently an open question.}
Consequently, there is a Gibbs distribution over their configurations.  Now,
there are many examples of hydrodynamic systems, strongly non-equilibrium in
their (standard) thermodynamics, where there are nonetheless important objects
which follows Gibbs distributions with various kinds of effective interaction
potentials \citep{Cross-Hohenberg}.  Perhaps the most striking case of this is
vortex lines in turbulent fluids --- see \citet{Chorin-vorticity} for a full
treatment.  For conventional statistical mechanics, this is just so much dumb
luck, but from our point of view, it indicates that the vortex lines (or other
coherent, structuring objects) are the local causal states.  Or rather: if what
we find doesn't look Gibbsian, it means we can do better.

\subsection{Building Coarse Grainings in Cellular Automata and Lattice
  Gases}

Cellular automata and lattice gases are fully-discretized classical field
theories.  That is, time is discrete, space is a discrete regular lattice, and
each point or ``cell'' can take one of a finite number of states at any one
time.  The state of each cell at time $t+1$ is a fixed, possibly stochastic,
function of the state of the cell at time $t$, along with the states of the
cells in a fixed ``neighborhood,'' thus preserving the nice classical property
of local interaction.  (All cells update in parallel.)  Originally introduced
to model mechanical self-reproduction
\citep{von-Neumann-self-reproducing,Poundstone-recursive}, cellular automata
have proved useful as models of many natural phenomena
\citep{Gutowitz-CA-conference,Chopard-Droz-text,Rothman-Zaleski-text}, as well
as mathematically fascinating objects in their own right
\citep{new-constructions-in-CA}.  They are important to us here because (1)
dynamic models of spin systems are stochastic CAs, and (2) they illustrate the
strategy we are advocating, at least as a matter of tacit practice.

When one simulates a CA, one knows, exactly, both the underlying microstate and
its dynamics.  It can nonetheless be very hard to say what it will do and why
it will do it.  This is not simply because some CA have high computational
complexity \citep{Burks-essays,Griffeath-Moore-LwoD,Moore-majority-vote}.
Rather, it is because the raw microstate is too detailed to be of use ---
$\Cmu$ is much too high.  One gains understanding by {\em deliberately}
throwing away most of the microscopic information, finding instead
coarse-grained observations where the dynamics are simpler to grasp
\citep{JPC-semantics}.  Generally, this means constructing macrostates with
well-behaved Markovian dynamics.  There are numerous examples of this strategy
in the literature, including the many derivations of hydrodynamics in lattice
gases \citep{Rothman-Zaleski-text}, the theory of heat conduction in the Creutz
CA \citep{Saito-Takesue-Miyashita}, or the vortex dynamics of the
zero-temperature Potts model \citep*{Moore-Nordahl-Minar-Shalizi}.  The goal,
always, is to throw away as much detail as possible, while retaining
information relevant to certain aspects of the large-scale dynamics --- to find
simple but accurate representations.  (Simple, inaccurate representations are
of course easy to find.)  Spatial computational mechanics
\citep*{Turbulent-pattern-bases,Hanson-thesis,Comp-mech-of-CA-example,%
  Hordijks-rule} provides tools whereby one can automatically find and filter
out low-information patterns, concentrating one's attention on higher-level,
information-rich emergent structures.

\section{What Is Not Being Said}

There are a number of puzzles about macrovariables which our arguments do {\em
not} resolve.

1. Why are so many good macrovariables extensive quantities?  It certainly does
not seem to follow from the fact that complete sets of macrovariables are
causal states.  We suspect, however, that something could be made of the
following sketch of an argument.  Extensive variables, by definition, add
across sub-systems.  If those sub-systems are independent, or nearly so, then
their totals will have large deviation properties
\citep{Ellis-entropy,Ellis-large-deviations-since-Boltzmann}.  In other words,
they will become increasingly well-behaved, statistically, in large systems.
This makes them good candidates for experimental observation.  Conceivably,
there are many extensive variables, other than the ones we commonly observe,
which, while subject to large-deviation principles for their additive
fluctuations, are ill-behaved (non-Markovian) over time, and so we ignore them.
Conversely, when approximate independence across sub-systems is violated, the
good macrovariables are non-extensive, and we need Tsallis statistics (rather
than the usual central-limit-theorem statistics).

2. Why are almost all good macrovariables derivatives of thermodynamic
potentials?  A deflating answer would be that candidate thermodynamic
potentials are under intense selection pressure for just this property.

3. Why are some good macrovariables reusable, e.g., why is temperature a good
macrovariable for almost everything?  It is not simply that we can only observe
a few variables and so observe them for almost everything, because order
parameters, for instance, are generally not reusable (no sense measuring the
nematic director in an antiferromagnet).  Moreover, why do these variables
typically take similar roles in systems with radically different microphysics?
(E.g., temperature again.)  To the best of our knowledge, no one has an answer
to these questions; we certainly don't.

Finally, our approach does not say why it is legitimate to treat a large,
locally unstable mechanical system stochastically in the first place.  We have
simply assumed that, since we are considering coarse-grained observations, it
is legitimate to deal with them statistically.  While the maximum entropy
principle provides a superficially attractive justification for this, it is
open to grave philosophical objections
\citep{Sklar-chance,Guttmann-prob-in-stat-phys}.  Worse,
\citet{Amari-hierarchical-info-geo} has shown that maximum entropy
distributions are simply those that minimize the degree of statistical
dependence between variables.\footnote{Intuitively, this makes sense.  The
  entropy, in bits, is the minimum mean number of binary variables needed to
  specify a sample drawn from the distribution.  Dependencies can be used to
  shorten the description, hence maximizing the entropy requires minimizing
  dependencies.  Showing this in detail requires an excursion through
  information geometry, and we refer the interested reader to Amari's paper.}
If distributions evolve towards minimal dependency, that is surely just a
contingent fact about the dynamics, rather than a universal principle of
inference.  We believe that the answer lies rather in ergodic theory
\citep{Khinchin-stat-mech,Mackey-times-arrow}, particularly recent developments
which emphasize the rapid mixing of low-dimensional projections of
high-dimensional smooth dynamics
\citep{Ruelle-smooth-dynamics,Dorfman-chaos-in-noneq,Gaspard-chaos-scattering-stat-mech},
plus the philosophical assumption that the initial conditions of the world are
``generic''.

\section{Conclusion}

Let us recap by way of telling a story.  Nameless men in black approach us, as
reputable practitioners of statistical mechanics, with a physical system, in
this case, a beaker full of gooey, shiny black stuff that sometimes moves
spontaneously.  We are able to probe certain aspects of it by physically
coupling to it --- e.g., we can X-ray it, take photographs, attach voltmeters,
scatter neutrons through it.  The men in black want us to predict certain
properties of the black oil, say, what will cause it to quiver in different
ways.
A mixture of interest and feasibility thus dictates an initial choice of
macrovariable.  Given this, we attempt to refine our predictive capabilities by
considering histories of observations.  From them we construct causal states.
If the causal states do not coincide with our initial observables, we either
supplement them with new variables, in a way which can be determined from the
causal states construction; or we eliminate unphysical distinctions and
unpredictive variables, again on the basis of the causal states.  When we need
supplementary variables, we can either devise new experimental methods to
observe them, probing new aspects of the physics, or we can merely construct
them logically, from histories of our original observables.  At the end of this
process we have the minimal set of variables from which we can optimally
predict the macrovariables of interest; ones which are, moreover, causally
complete.

Our initial choice of macrovariables is the product of our ability to observe
the system, and our choices about what to predict.  Beyond that initial choice,
and the requirement that good macrostates have certain causal properties, the
causal state we use are completely out of our control, fixed entirely by the
objective, microphysical dynamics.  A different set of initial variables will,
generally, lead to a different set of causal states.  Sometimes, but not
always, these causal states are related in a hierarchy of emergence.  One might
put it like this: for every question we ask It, Nature has a definite answer;
but Nature has no preferred questions.

\section*{Acknowledgments}

We thank David Albert, Sven Brueckner, Jim Crutchfield, Michael Lachmann, Erik
van Nimwegen, Scott Page, Jay Palmer and Peter Stadler for helpful discussions.
CRS thanks Prof.\ Crutchfield for generous support and encouragement, Kara Kedi
for providing a reason to get out of bed each morning, and Kristina Shalizi for
being perfect.  His work was supported at SFI by the Dynamics of Learning
project, under DARPA contractual agreement F30602-00-2-0583, and at the
University of Michigan by grant from the James S. McDonnell Foundation.  CM
thanks Itamar Pitowsky for organizing the International Conference on the
Conceptual Foundations of Statistical Mechanics in Jerusalem in May 2000, and
Michel Morvan and the Laboratoire d'Informatique Algorithmique Fondements et
Applications (LIAFA) at Universit\'e Paris 7 for a visit during which some of
this paper was written.  He is supported by NSF grant PHY-0071139, the Sandia
University Research Program, and Los Alamos National Laboratory.  He also
thanks Tracy Conrad and Spootie the Cat for fleeting glimpses of enlightenment

\appendix

\section{Information Theory}
\label{app:info-theory}

The information contained in a discrete random variable $X$, also called its
{\em entropy} or {\em Shannon entropy}, is
\begin{eqnarray*}
H[X] & \equiv & -\sum_{x}{\Prob(X=x)\log_2{\Prob(X=x)}}\\
& = & -\left\langle\log_2{\Prob(X)}\right\rangle
\end{eqnarray*}
It is the smallest number of bits (binary distinctions) needed, on average, to
specify the value of $X$.  We may think of it as the uncertainty an ideal
observer, who knew the true ensemble $X$ is drawn from, would have about $X$.

The {\em joint entropy} of two variables $X$ and $Y$ is defined similarly,
\[
H[X,Y] \;\equiv\; -\sum_{x,y}{\Prob(X=x,Y=y)\log_2{\Prob(X=x,Y=y)}}
\]
It is easy to show that $H[X,Y] \leq H[X] + H[Y]$, with equality if and only if
$X$ and $Y$ are statistically independent.

The {\em conditional entropy} of $X$ given $Y$, is
\begin{eqnarray*}
H[X|Y] & \equiv & H[X,Y] - H[Y]\\
& = & -\sum_{y}{\Prob(Y=y)\sum_{x}{\Prob(X=x|Y=y)\log_2{\Prob(X=x|Y=y)}}}\\
& = & \sum_{y}{\Prob(Y=y) H[X|Y=y]}
\end{eqnarray*}
Then $H[X|Y=y]$ is the information needed to specify $X$ in the sub-ensemble
where $Y$ has the value $y$, and $H[X|Y]$ is the average information remaining
in $X$ given $Y$.

Finally, the {\em mutual information} between $X$ and $Y$ is
\begin{eqnarray*}
I[X;Y] & \equiv & H[X] + H[Y] - H[X,Y]\\
& = & H[X] - H[X|Y] \\
& = & H[Y] - H[Y|X] \;\leq\; H[Y] \enspace ,
\end{eqnarray*}
that is, the amount by which knowledge of one variable reduces the uncertainty
in the other.

\section{Markov processes}
\label{app:markov}

Suppose a process generates a probability distribution over time series,
$\ldots,s_{t-1},s_t,s_{t+1},\ldots$.  It is a (first-order) {\em Markov
  process} if
\[ \Prob(s_{t+1} = s | \pastt = \ldots,s_{t-1},s_t) \;=\;
   \Prob(s_{t+1} = s | s_t) \enspace .
\]
In other words, the only dependence that $s_{t+1}$ has on its entire past
history is on its current state $s_t$, and previous values yield no additional
information about its future.  Examples of Markov processes include:
\begin{itemize}
\item A deterministic dynamical system where $s_{t+1} = f(s_t)$ for some
  function $f$
\item A series of fair coin flips, where $\Prob(s_{t+1} = s) = 1/2$ for $s \in
  \{{\rm heads},{\rm tails}\}$
\item Brownian motion, where $\Prob(s_{t+1} = x | s_t = y) = f(|x-y|)$ for a
  Gaussian function $f$
\end{itemize}
As an example of a non-Markovian process, suppose $s_t$ is the position at time
$t$ of a particle which is moving at constant velocity.  Here $s_{t+1}$ depends
on $s_t$ and $s_{t-1}$, i.e.\ its dynamics is second-order, so there are
correlations with the past that are not captured by the current state.  On the
other hand, if we expand our set of observables so that $s_t$ includes both the
particle's position and its velocity, then the process becomes first-order
Markovian.

\section{Partitions}

A {\em partition} $P$ of a set $\Omega$ is a set $P_i$ of subsets of $\Omega$
which are mutually exclusive and jointly exhaustive.  That is, $P_i \cap P_j =
\emptyset$ (unless $i = j$), and $\Omega = \bigcup_{i}{P_i}$.  The sets in $P$
are the {\em cells} of the partition.

An {\em equivalence relation} or {\em equivalence} $\sim$ on $\Omega$ is a
relation which is reflexive, symmetric and transitive: $a \sim a$, $(a \sim b)
\Leftrightarrow (b \sim a)$ and $(a \sim b) \wedge (b \sim c) \Rightarrow (a
\sim c)$.  The {\em equivalence class} of a point is the set of all points
which are equivalent to it.  We write the equivalence class of $x$ as $[x]$;
$[x] = \left\{y|x\sim y\right\}$.  Since every point is equivalent to itself,
every point has a non-empty equivalence class, and every point belongs to some
equivalence class.

\begin{proposition}
Every partition corresponds to an equivalence relation, and vice versa.
Equivalence classes are cells of the partition.
\end{proposition}

{\em Proof}.  First, we construct an equivalence relation from a partition.
Simply say that $x \sim y$ iff $x$ and $y$ are in the same cell.  This is
symmetric, reflexive and transitive, hence an equivalence.  Now we build a
partition from an equivalence relation.  We claim that the equivalence classes
are mutually exclusive and jointly exhaustive.  Mutually exclusive means that
either $[x] = [y]$ or $[x] \cap [y] = 0$, for all $x$ and $y$.  To see this,
consider any point $z \sim x$.  Now, $y \sim z$ if and only if $y \sim x$ ---
if by transitivity, and only if likewise.  Hence $x \sim y$ iff $[x] = [y]$.
If $x \not \sim y$, then there cannot exist even one $z$ such that $z \sim x$,
and so no point belongs to the intersection of $[x]$ and $[y]$.  Since every
point has an equivalence class, the set of equivalence classes is exhaustive.
QED.

Every function $f$ on $\Omega$ induces an equivalence relation $\sim_f$, thus:
$a \sim_f b$ iff $f(a) = f(b)$.  Similarly every equivalence relation defines
an (infinite) class of functions: give each equivalence class a unique label
and map points to their equivalence-class labels.  Hence every function defines
a partition and vice versa.

The {\em identity} partition is one where each cell contains only a single
element of $\Omega$, i.e., where each equivalence class consists of a single
point.  The {\em trivial} partition is the one which contains only a single
cell, equal to $\Omega$.

One partition, $A$, is {\em finer} than another, $B$, iff for each $a \in A$,
there exists a $b \in B$ such that $a \subseteq b$, and, for at least one $a$,
$a \subset b$.  Then $A$ is a {\em refinement} of $B$, and $B$ is {\em coarser}
than $A$. Refinement always increases cardinality.  If neither $A$ nor $B$ is a
refinement of one another, and they are not equal, they are {\em incomparable}.

Let $A$ and $B$ be two partitions.  Construct all the sets formed by taking the
intersection of one cell from $A$ with one cell from $B$.  This collection is
also a partition, the {\em product} of $A$ and $B$.  Symbolically, $A\cdot B =
\left\{c|\exists a \in A, b \in b, c = a \cap b\right\}$.  It is a refinement
of both $A$ and $B$.

\begin{proposition}[Factoring Refinements]
\label{proposition:factoring-refinements}
Let $A$ be a partition and $B$ be any refinement of $A$.  Then there exists a
{\em minimal factor partition} $C$ such that $B = A\cdot C$ and $B \neq A\cdot
D$ for any $D$ with fewer cells than $C$.  If $A$ and $B$ are both finite, then
the minimal factor partitions are themselves finite, and there is a finite
number of them.
\end{proposition}

{\em Proof}.  We construct a minimal factor $C$.  Each cell $a$ of $A$ contains
$n_a$ cells from $B$; $a = \bigcup_{i=1}^{n_a}{b^{a}_{i}}$.  Let $N$ be the
maximum of $n_a$ over all the cells of $A$.  Define $c_i = \bigcup_{a\in
  A}{b^{a}_{i}}$.  Clearly $C = \left\{c_i\right\}$ is a partition, and equally
clearly its product with $A$ will be $B$.  Any partition whose product with $A$
is $B$ must have a cardinality of at least $N$, because it must break (at
least) one cell of $A$ into $N$ sub-cells.  Hence we have constructed a minimal
$C$ whose product with $A$ gives the desired refinement.  Moreover any $D$ such
that $B = A\cdot D$ must be a refinement of some minimal factor $C$.  The
number of minimal factors is at most the number of ways of labeling the
subcells of $A$, viz., $\prod_{a\in A}{n_a!}$.

Note that minimal factors are incomparable to $A$.  They are not refinements,
because each cell of the factor is not entirely contained within a single cell
of $A$.  But conversely, there are cells in $A$ which are not entirely
contained within a single cell of the factor.  Obviously, they are not equal to
$A$.  Hence they are incomparable.

\bibliography{locusts}
\bibliographystyle{crs}

\end{document}